\documentclass[pre,aps,twocolumn,showpacs]{revtex4}

\usepackage{graphicx}
\graphicspath{{Figures/},{Figures/hiRes/}}      %%% Low resolution if available
\DeclareGraphicsExtensions{.eps,.ps}
\usepackage{amsmath}

\newcommand\Ef{E_F}

\begin{document}

\title{Front propagation into unstable metal nanowires}

\author{J.\ B\"urki}
\email{buerki@physics.arizona.edu}

\affiliation{Department of Physics, University of Arizona, Tucson, AZ 85721}

%\date{\today}
\date{April 17, 2007}

\begin{abstract}
Long, cylindrical metal nanowires have recently been observed to form and be stable for seconds at a time at room temperature.
Their stability and structural dynamics is well described by a continuum model, the nanoscale free-electron model, which predicts cylinders in certain intervals of radius to be linearly unstable.
In this paper, I study how a small, localized perturbation of such an unstable wire grows exponentially and propagates along the wire with a well-defined front.
The front is found to be pulled, and forms a coherent pattern behind it.
It is well described by a linear marginal stability analysis of front propagation into an unstable state.
In some cases, nonlinearities of the wire dynamics are found to trigger an invasive mode that pushes the front.
Experimental procedures that could lead to the observation of this phenomenon are suggested.
\end{abstract}


\maketitle \vskip2pc

%%%%%%%%%%%%%%%%%%%%%%%%%%%%%%%%%%%%%%%%%%%%%%%%%%%%%%%%%%%%%%%%%%%%%%%
%\section{Introduction}
\label{sec:intro}
%%%%%%%%%%%%%%%%%%%%%%%%%%%%%%%%%%%%%%%%%%%%%%%%%%%%%%%%%%%%%%%%%%%%%%%

Front propagation into unstable states occurs in many areas of physics, chemistry, and biology (see Ref.\ \cite{vansaarloos03} for a recent review), and is often related to pattern formation mechanisms.
In this paper, I show that metal nanowires, whose dynamics can be described by a continuum model, the nanoscale free-electron model (NFEM) \cite{stafford97,burki03,burki05}, exhibit such front propagation into unstable cylinders.

Recent transmission electron microscopy experiments \cite{kondo97,kondo00,rodrigues00,rodrigues02} have observed gold and silver nanowires to form long cylinders, with diameters of order one nanometer, that are stable for seconds at a time at room temperature.
A theoretical description within the NFEM shows \cite{burki03,burki04} that such self-assembly is natural if conditions are such that the atom mobility allows the wire to explore its configuration space and reach its equilibrium shape, as is the case at room temperature for these metals.

% Gold and silver nanowires have been shown recently, both experimentally using transmission electron microscopy \cite{kondo97,kondo00,rodrigues00,rodrigues02}, and theoretically within the NFEM \cite{burki03,burki04}, to self-assemble into long cylinders, with diameters of order one nanometer, provided  conditions are such that the atom mobility allows the wire to explore its configuration space and reach its equilibrium shape, as is the case at room temperature for these metals.

Due to the high number of surface atoms---with low coordination numbers---in such nanowires, surface effects are particularly important, and favor wire break-up due to the Rayleigh instability \cite{chandrasekhar81,kassubek01}.
It has been shown, using the NFEM \cite{zhang03,burki05}, that the quantum confinement of electrons in the cross-section of the wire provide electron-shell effects---similar to those well-known in cluster physics \cite{brack93}---that compete with surface effects, and stabilize cylindrical wires for a finite range around magic radii \cite{kassubek01,zhang03}, as well as a number of number of wires with broken axial symmetry \cite{urban04,urban06}.

The NFEM \cite{stafford97,burki03,burki05} is a continuum model where the atomic structure is replaced by a uniform, positively charged background, and the emphasis is put on the electronic structure.
An extension of the model \cite{burki03} includes ionic dynamics through surface self-diffusion, which is expected to dominate the structural dynamics in such thin wires.
One of its important predictions is that a random wire naturally evolves into a universal equilibrium shape consisting of a perfect cylinder of a magic radius, connected to thicker leads \cite{burki03,burki04}.
Recently, the rich kink dynamics of the NFEM has been described \cite{burki07},  and shown to be qualitatively similar to the thinning mechanism of gold nanowires observed experimentally \cite{oshima03}, proving it to be a suitable, if simplified, model of the structural dynamics of metal nanowires.

This paper is concerned with radii outside of the intervals of stability, and studies their dynamics under surface self-diffusion. 
A localized perturbation to an unstable cylinder is found to grow exponentially, and to propagate along the unstable wire with a well defined front, which is a coherent pattern forming front, and can be either pulled or pushed depending on the wire radius.

The paper is organized as follows: In Sec.\ \ref{sec:model}, the NFEM is introduced and some of its main results directly relevant to the present article are summarized. 
Numerical simulations of the dynamics of an unstable cylinder are presented in Sec.\ \ref{sec:instabprop}, while the front propagation analysis is developed in Sec.\ \ref{sec:frontprop}.
Section\ \ref{sec:concl} discusses the results and some experimental setups that could detect front propagation into unstable metal nanowires.

%%%%%%%%%%%%%%%%%%%%%%%%%%%%%%%%%%%%%%%%%%%%%%%%%%%%%%%%%%%%%%%%%%%%%%%
\section{The nanoscale free-electron model}\label{sec:model}
%%%%%%%%%%%%%%%%%%%%%%%%%%%%%%%%%%%%%%%%%%%%%%%%%%%%%%%%%%%%%%%%%%%%%%%

The NFEM is a continuum model of open metallic nanosystems with an emphasis on the electronic structure, which is treated exactly \cite{stafford97,burki05}.
It is thus particularly suitable as a model of metal nanowires, and successfully describes many of their equilibrium \cite{stafford97,kassubek99,kassubek01,zhang03,urban03,urban06} and dynamical \cite{burki03,burki05a,burki04,burki05,burki07} properties in simple physical terms.

The discrete atomic structure is replaced by a uniform, positively-charged background (Jellium), which provides a confining potential for electrons. 
Electronic degrees of freedom are described using a free-electron model, thus neglecting interactions, except inasmuch as they rescale macroscopic quantities such as the bulk energy density $\omega_B$, and the surface tension $\sigma_s$ \cite{kassubek99,zhang03,burki05}.
This model is particularly suitable for simple metals, with good screening and a close-to-spherical Fermi surface, such as those with a single $s$-electron conduction band at the Fermi surface.
Such conditions are fulfilled for alkali metals like sodium, and for noble metals such as gold and silver, although $d$ electrons may play a role for noble metals.

While the NFEM has no such restriction, only axisymmetric wires are considered in this paper.
This choice greatly simplifies the numerical treatment of the dynamics, as the wire shape can be described by a single radius function $R(z,t)$.
It is justified by the facts that 
({\sl i}) the most stable wires are axisymmetric \cite{urban04,urban06}, and ({\sl ii}) the dynamics tends to decrease surface area, and thus further favors axisymmetric wires.

A nanowire being an open system, the electronic energy is given by the grand canonical potential $\Omega_e$.
Like any extensive thermodynamic quantity, $\Omega_e$ can be written as a Weyl expansion \cite{brack97}---a series in geometrical quantities such as system volume ${\cal V}$ and surface area ${\cal S}$---complemented by a mesoscopic, fluctuating contribution $\delta\Omega$:
 \begin{equation}\label{eq:Omega_e}
  \Omega_e[R(z)] = \omega_B{\cal V} + \sigma_s{\cal S} +\delta\Omega,
 \end{equation}
where the values of $\omega_B$ and $\sigma_S$ may be chosen to match the bulk properties of the metal to be described.
As results are independent of $\omega_B$, its free-electron value $\omega_B=-2\Ef k_F^3/15\pi^2$ is used, while the surface tension is set to $\sigma_s=1.256\,$N/m, a value appropriate for the description of gold \cite{tyson77}.

\begin{figure}[b]
  \includegraphics[width=0.99\columnwidth]{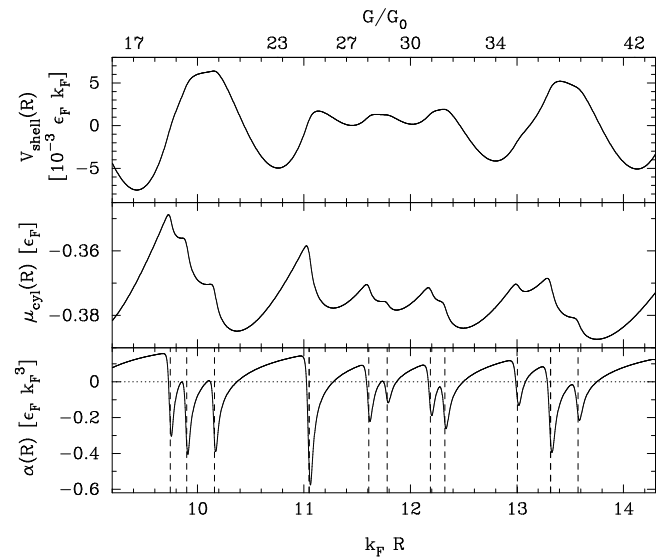}
  \caption{Electron-shell potential $V_{shell}(R)$ (top), chemical potential $\mu_{cyl}(R)$ (middle), 
        and stability coefficient $\alpha(R)$ (bottom) for a cylindrical wire,
        as a function of its dimensionless radius $k_FR$, where $k_F$ is the Fermi wavevector.
        (The radius range is limited to that corresponding to the simulations presented here. 
        For a graph over a more extended range, see Refs.\ \onlinecite{burki05,kassubek01,burki03}.)
        Vertical dashed lines in the bottom panel mark the positions of conductance channel openings, 
        which drive the instability \cite{kassubek01}.
  	The top axis shows the conductance values of linearly stable cylinders in units of the 
  	conductance quantum, $G_0=2e^2/h$.
  }\label{fig:Vshell}
\end{figure}

Assuming the wire cross section varies slowly along the wire (adiabatic approximation), the mesoscopic contribution may be written as 
\begin{equation}\label{eq:dOmega}
 \delta\Omega[R(z)] = \int_0^L\!dz\,V_{shell}[R(z)],
\end{equation}
where the electron-shell potential $V_{shell}(R)$ can be computed using a semi-classical approximation \cite{burki05}.
$V_{shell}(R)$, depicted in Fig.\ \ref{fig:Vshell} (top panel) for a cylindrical wire as a function of its radius $R$, is responsible for stabilizing wires of magic radii, which correspond to its deep minima.

The ionic dynamics is taken to be classical and can be assumed to occur mainly through surface self-diffusion, as most atoms in thin metal wires are surface atoms \cite{SurfAtoms,burki03}.
The evolution equation for the radius function $R(z,t)$ derives from ionic mass conservation
\begin{equation}\label{eq:diffusion}  %%%  Eq. 3  %%%%%%%%%%%%%%%%%%%%%
%%%%%%%%%%%%%%%%%%%%%%%%%%%%%%%%%%%%%%%%%%%%%%%%%%%%%%%%%%%%%%%%%%%%%%%
  \frac{\pi}{{\cal V}_a}\frac{\partial R^2(z,t)}{\partial t}
        + \frac{\partial}{\partial z}J_z(z,t)
%	+ \frac{\partial}{\partial z}\bigl[2\pi R(z,t)J_z(z,t)\bigr]
	= 0,
\end{equation}  %%%%%%%%%%%%%%%%%%%%%%%%%%%%%%%%%%%%%%%%%%%%%%%%%%%%%%%
where ${\cal V}_a=3\pi^2/k_F^3$ is the volume of an atom, and the $z$-component $J_z$ of the total surface current is given by Fick's law:
\begin{equation}\label{eq:current} %%%  Eq. 4  %%%%%%%%%%%%%%%%%%%%%%%%
%%%%%%%%%%%%%%%%%%%%%%%%%%%%%%%%%%%%%%%%%%%%%%%%%%%%%%%%%%%%%%%%%%%%%%%
  J_z = -\frac{\rho_SD_S}{k_B T}\frac{2\pi R(z,t)}{\sqrt{1+(\partial_zR)^2}}
        \frac{\partial\mu}{\partial z}.
\end{equation}  %%%%%%%%%%%%%%%%%%%%%%%%%%%%%%%%%%%%%%%%%%%%%%%%%%%%%%%
Here $\rho_S$ and $D_S$ are, respectively, the surface density of atoms and
the surface self-diffusion coefficient, and $\partial_zR=\partial R/\partial z$.

The precise value of $D_S$ is not known for most metals, but it can be removed from the evolution equation by rescaling time to the dimensionless variable $\tau=\omega_0t$, with the characteristic temperature-dependent frequency $\omega_0=\rho_SD_ST_F/T$.
For comparison to experimental time scales, one can estimate that for quasi-one-dimensional diffusion $D_s \approx \nu_D a^2 \exp(-E_s/k_B T)$,
where $\nu_D$ is the Debye frequency, $a$ is the lattice spacing, and $E_s$ is an
activation energy comparable to the energy of a single bond in the solid.

The chemical potential $\mu[R(z)]$ of a surface atom can be computed from the energy change due to the local addition of the volume ${\cal V}_a$ of an atom to the system.
Within the Born-Oppenheimer approximation, and assuming the electrons act as an incompressible fluid \cite{zhang03,burki05,incompFluid}, the chemical potential is given \cite{burki03} by the functional derivative $\mu[R(z)]={\cal V}_a/(2\pi R)\cdot\delta\Omega_e/\delta R(z)$. 
Starting from Eqs. (\ref{eq:Omega_e}) and (\ref{eq:dOmega}), one obtains
\begin{equation}\label{eq:mu}  %%%  Eq. 5  %%%%%%%%%%%%%%%%%%%%%%%%%%%%
%%%%%%%%%%%%%%%%%%%%%%%%%%%%%%%%%%%%%%%%%%%%%%%%%%%%%%%%%%%%%%%%%%%%%%%
  \mu[R(z)] = \mu_0
    + \frac{{\cal V}_a}{2\pi R}\left(
      \frac{2\sigma_s\partial{\cal C}[R(z)]}{\sqrt{1+(\partial_z R)^2}}
      + \frac{\partial V_{shell}}{\partial R}\right),
\end{equation}  %%%%%%%%%%%%%%%%%%%%%%%%%%%%%%%%%%%%%%%%%%%%%%%%%%%%%%%
where $\mu_0=\omega_B {\cal V}_a$ is the bulk chemical potential.
Here $\partial{\cal C}[R(z)] = \pi\left(1 - \frac{R \, \partial^2_{z}R}{1+(\partial_zR)^2}\right)$ is the local mean curvature of the wire, and arises from the functional derivative of the surface term in Eq.\ (\ref{eq:Omega_e}).
The chemical potential of a cylinder $\mu_{cyl}(R)\equiv\mu[R(z)=R]$ is plotted as a function of the radius $R$ in the middle panel of Fig.\ \ref{fig:Vshell}.

The appropriate boundary conditions when simulating a cylinder, which in a real system is connected to larger leads \cite{burki03,burki04}, have been shown to be Neumann boundary conditions, $\partial_z R = 0$ at both wire ends \cite{burki03,burki05a}.

Note that, despite the apparent simplicity of Eqs.\ (\ref{eq:diffusion})--(\ref{eq:mu}) as written above, the resulting partial differential equation for $R(z,t)$ is fourth-order in $z$-derivatives and highly nonlinear.
Its classical counterpart, with $V_{shell}\equiv 0$, has been extensively studied \cite{coleman96,bernoff98,eggers98}.
It was shown to have stationary states corresponding to shapes of constant mean curvature: The sphere, the cylinder and the unduloid of revolution, the latter being always unstable.
The addition of the electron-shell-potential term, of quantum-mechanical origin, is responsible for a rich dynamics, discussed extensively in Ref.\ \cite{burki07} and in this paper, as well as for the stabilization of the unduloid as a stationary state \cite{burki03,burki04}.
Without this term, any cylinder longer that its perimeter is unstable toward the Rayleigh instability and breaks apart into spheres. 
In that case, the system exhibits front propagation \cite{powers98}, but there is no saturation of the instability, and the radius dependence of the dynamics is trivial. Furthermore, nonlinear effects do not seem to influence the front propagation.

\subsection*{Summary: Linear stability analysis}  %{Summary of previous results}
%%%%%%%%%%%%%%%%%%%%%%%%%%%%%%%%%%%%%%%%%%%%%%%%%%%%%%%%%%%%%%%%

A linear stability analysis of cylindrical wires has been performed within the NFEM \cite{kassubek01,zhang03,urban03,burki05}.
The change in the energy (\ref{eq:Omega_e}) due to a radius perturbation $\Delta R = \sum_qb_q\exp(i q z)$ was found to be,
\begin{equation}\label{eq:dOmega_e}
  \Delta\Omega_e = L\sum_q\alpha(R_0;q)|b_q|^2,
\end{equation}
so that the sign of $\alpha(R_0;q)$ determines the linear stability of a wire of radius $R_0$ towards a perturbation of wavevector $q$. 
Hence $\alpha(R_0;q)$ has been named the {\sl stability coefficient}. 
It was further found that the global linear stability is essentially determined by the long wavelength limit \cite{kassubek01,zhang03,urban03} $\alpha(R_0)\equiv\alpha(R_0;q=0)$, with
\begin{equation}\label{eq:alpha}
  \alpha(R_0)=\left(-\frac{2\pi\sigma_S}{R}+\left.\frac{d^2V_{shell}}{dR^2}-\frac1{R}\frac{dV_{shell}}{dR}\right)\right|_{R=R_0}.
\end{equation}

Instabilities were found to result from a transverse eigenenergy of the wire crossing the Fermi energy $\Ef$, thus closing or opening a conduction channel \cite{stafford97,kassubek01}.
These thresholds are marked by vertical dashed lines in the plot of $\alpha(R)$ in the bottom panel of Fig. \ref{fig:Vshell}.
Cylinders with radii in the vicinity of ``magic'' radii---corresponding to minima of the shell potential $V_{shell}$, see Fig.\ \ref{fig:Vshell}---are linearly stable, while wires close to maxima of $V_{shell}$ are unstable.

A linearized dynamical theory \cite{zhang03} shows that unstable wires develop an exponentially growing instability with a well-defined wavelength $\lambda=2\pi/q_{max}$, corresponding to the maximally unstable mode, such that $\alpha(R_0;q_{max})$ is extremal.
This instability was argued to saturate, and eventually lead to a phase separation of the wire into thick and thin segments of stable radii \cite{zhang03}.
Simulations using the full dynamics Eqs.\ (\ref{eq:diffusion})--(\ref{eq:mu}) have confirmed this \cite{burki03}, and shown that the phase separation occurs via a complex dynamics involving kink interactions and annihilation \cite{burki07}.
In this paper, I analyze the growth and propagation of such instabilities in greater detail.

%%%%%%%%%%%%%%%%%%%%%%%%%%%%%%%%%%%%%%%%%%%%%%%%%%%%%%%%%%%%%%%%%%%%%%%
\section{Instability propagation}\label{sec:instabprop}
%%%%%%%%%%%%%%%%%%%%%%%%%%%%%%%%%%%%%%%%%%%%%%%%%%%%%%%%%%%%%%%%%%%%%%%

Results on the evolution of an unstable cylinder, including the growth of a single-wavelength perturbation followed by  phase separation, were briefly presented in a Letter \cite{burki03}.
In this section, I provide more details about simulations of the growth and propagation of perturbed linearly unstable wires as a function of the wire radius $R_0$.
An initially localized perturbation is found to grow exponentially and propagate with a well-defined front, which is analyzed in detail.
A theoretical analysis of the front propagation is provided in Sec.\ \ref{sec:frontprop}.

\begin{figure}[bh]
 \centering
 \includegraphics[width=0.99\columnwidth]{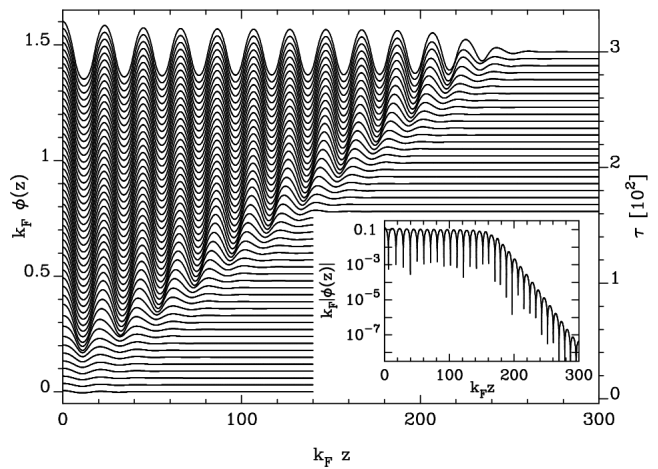}
 \caption{Instability propagation for a wire of radius $k_FR_0=11.65$: 
          The main panel shows $\phi(z)=R(z)-R_0$ at equidistant times (shifted vertically for clarity), 
          illustrating a typical time evolution of the instability propagation. 
          In particular, the coherent pattern left behind the front can be observed.
          The r.h.s.\ axis gives the dimensionless evolution time $\tau$ for the corresponding curves.
          The inset shows $|\phi(z)|$ at a given time, $\tau\sim180$, on a logarithmic scale, illustrating 
          its exponential decay ahead of the front.
 }\label{fig:frontprop}
\end{figure}

The initial condition for these simulations is an unstable cylinder of radius $R_0$, i.e.\ a radius such that $\alpha(R_0)<0$.
Two lengths, $k_FL=300$ and $400$, have been considered.
A Gaussian perturbation of amplitude $k_F\delta R = 0.01$, localized at the left boundary, $z=0$, is added to trigger the instability.
Results have been checked to be independent of the type, amplitude, and extension of the initial perturbation, as long as it is sufficiently localized \cite{vansaarloos03}.

After a brief initial incubation period, during which the amplitude of the perturbation decreases, it grows exponentially and propagates along the wire.
A well-defined front forms and moves at a constant velocity, as illustrated in Fig.\ \ref{fig:frontprop}, where the wire perturbation $\phi(z,\tau)=R(z,\tau)-R_0$ for a wire with $k_FR_0=11.65$, is plotted at equidistant times $\tau$, with a vertical shift proportional to $\tau$.
Behind the front, a coherent pattern consisting of a single-wavelength radius oscillation forms and saturates at an amplitude $k_F\Delta\phi\sim0.1$--$1$, whose value depends on the details of the shell potential $V_{shell}$ (cf.\ Fig.\ \ref{fig:Vshell}) in the vicinity of the unstable radius $R_0$ considered.

A logarithmic plot of (the absolute value of) the wire perturbation $|\phi(z,t)|$ (see inset of Fig.\ \ref{fig:frontprop}) clearly shows its exponential decay ahead of the front, whose position $z_f$ is defined as the point where the perturbation amplitude reaches a certain threshold value $c$ \cite{threshold}.
Tracking $z_f$ as a function of time, the front velocity $v$ can be extracted, and is found to be constant after an initial period when it is influenced by boundary and initial conditions.
Similarly, the wavevector $q$ and decay length $\xi$ of the perturbation ahead of the front can be extracted.
Both are found to be constant after an initial decrease during early evolution.

In the simpler cases, further evolution of the wire behind the front occurs on a timescale orders of magnitude longer than the front propagation.
It involves annihilation of kink/antikink pairs---upward and downward steps in $R(z)$, whose dynamics has been described in Ref.\ \cite{burki07}---and eventually leads to phase separation into thick and thin segments of stable cylinders \cite{zhang03,burki03}.

\begin{figure}[b]
  \includegraphics[width=0.95\columnwidth]{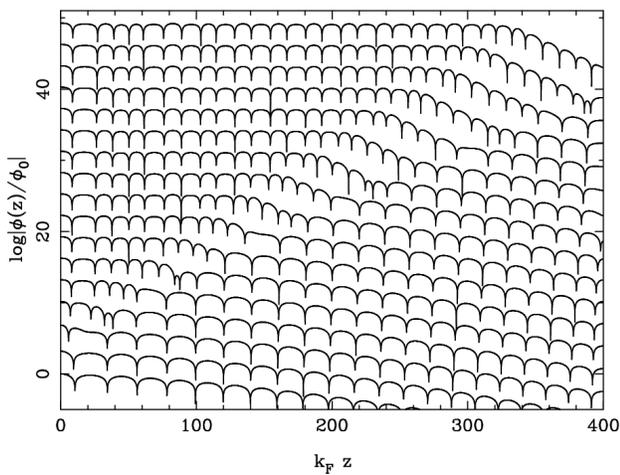}
  \caption{$\log|\phi(z,\tau)/\phi_0|$ at equidistant times $\tau$, shifted vertically for clarity, 
           for a wire of radius $k_FR=12.28$, showing an invasive mode. All curves are normalized by $\phi_0\equiv\phi(z=0,\tau=\tau_0)$, where $\tau_0$ is the time of the bottom curve.}
           \label{fig:dbledecay}
\end{figure}

In other cases however, nonlinear effects quickly take over the dynamics behind the front, and start pushing it.
This modifies its velocity, as well as the decay rate, and possibly the wavelength ahead of the front.
An example of such behavior is shown in Fig. \ref{fig:dbledecay}, where $\log|\phi(z,\tau)|$ is plotted at equidistant times (each curves shifted vertically) for a wire with radius $k_FR=12.28$.
One can clearly see a second front with a shorter decay length that progressively takes over the initial front.
Such a front will be referred to as a {\sl nonlinear} front, while the initial, slower front will be called {\sl linear}.

The velocity, decay length and wavevector of both fronts can be extracted from the simulations.
The three quantities are plotted as open circles in Fig.\  \ref{fig:lintheory} for the linear front, while filled light (green) circles are used for the nonlinear front, when it is detected.
All three quantities are found to depend non-trivially on the wire radius $R_0$.
They reach local extrema simultaneously for values of $R_0$ that correspond to minima of the stability coefficient $\alpha$ (Fig.\ \ref{fig:Vshell} and Eq.\ (\ref{eq:alpha})), i.e., maximally unstable wires.
The front velocity $v$ and wavevector $q$ vanish, while the decay length $\xi$ diverges, at the stability boundary $R_c$, where $\alpha(R_c)=0$.
Roughly, it seems that a nonlinear front appears for wires with $\alpha(R_0)\gtrsim 0.1\Ef k_F^3$.

%%%%%%%%%%%%%%%%%%%%%%%%%%%%%%%%%%%%%%%%%%%%%%%%%%%%%%%%%%%%%%%%%%%%%%%
\section{Front propagation analysis}\label{sec:frontprop}
%%%%%%%%%%%%%%%%%%%%%%%%%%%%%%%%%%%%%%%%%%%%%%%%%%%%%%%%%%%%%%%%%%%%%%%

\begin{figure}[tb]
 \centering
 \includegraphics[width=0.95\columnwidth]{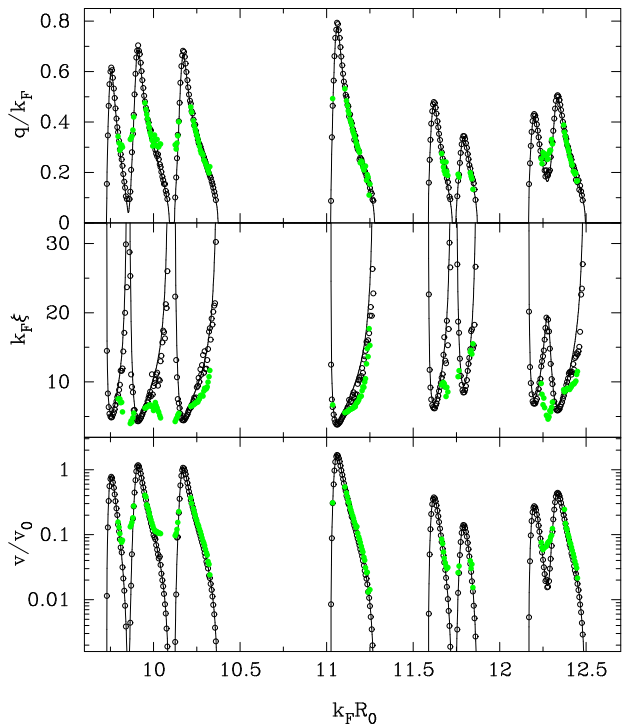}
 \caption{(color online) Instability wavevector $q$ (top), decay length $\xi$ (middle), 
          and front velocity $v$ (bottom) as a function of unstable wire radius $R_0$. 
          The solid lines give the results of the linear theory, Eqs.\ (\ref{eq:kstar}) 
          and (\ref{eq:Vfront}), and the circles are results extracted from the full 
          dynamical simulations (Error bars are smaller than the symbols, and therefore not plotted.)
          Open circles are used for the linear front, while filled (green) circles correspond to 
          nonlinear fronts.
          The velocity in the bottom panel is given in units of $v_0=\lambda_F\omega_0$.
 }\label{fig:lintheory}
\end{figure}

In this section, I provide a more detailed analysis of the instability propagation, showing in particular that  the evolution of the linear fronts---marked by open circles in Fig.\ \ref{fig:lintheory}---derives from the linearized dynamics, hence the name, and is a typical example of front propagation into an unstable state.
Following ideas from the linear marginal stability analysis of Refs.  \cite{vansaarloos88, vansaarloos03}, much of the front dynamics can be understood from the linearized evolution equation for the perturbation $\phi(z,t)$, which is 
\begin{equation}\label{eq:LinearEq}
 \frac{2\pi R_0}{{\cal V}_a}\frac{\partial\phi}{\partial t} =
  - \frac{\omega_0{\cal V}_a}{\Ef}
    \left[2\pi\sigma_sR_0\frac{\partial^4\phi}{\partial z^4} - \alpha(R_0)\frac{\partial^2\phi}{\partial z^2}
  \right].
\end{equation}
Assuming a front of the form $\phi(z,t) = \exp[i(\omega t-k z)]$, with $\omega$ and $k$ complex quantities, its dispersion relation
\begin{equation}\label{eq:omega}
 \omega(k) = i\frac{\omega_0\sigma_s{\cal V}_a^2}{\Ef} k^2
   \left(k^2+\frac{\alpha(R_0)}{2\pi\sigma_sR_0}\right)
\end{equation}
is derived.
It has been argued \cite{vansaarloos88} that the front wavevector $k^*$ is such that the front is marginally stable, and thus satisfies the conditions \cite{vansaarloos03}
\begin{equation}\label{eq:frontsatbility}
  \text{Im}\left.\frac{d\omega}{dk}\right|_{k^*} = 0,\qquad 
  \frac{\text{Im}\,\omega(k^*)}{\text{Im}\,k^*} = \left.\frac{d\omega}{dk}\right|_{k^*}.
\end{equation}
This yields the front wavevector $k^*\equiv q - i/\xi$,
\begin{equation}\label{eq:kstar}
 k^*=\frac{1}{4}\sqrt{-\frac{\alpha(R_0)}{\pi\sigma_sR_0}}
   \left(\sqrt{3+\sqrt7} - i\sqrt{\frac{\sqrt7-1}{3}}\right),
\end{equation}
where the real part $q$ determines the wavevector of the pattern left behind the front, and
the inverse $\xi$ of the imaginary part corresponds to the decay length ahead of the front.
They are respectively plotted as solid lines in Fig. \ref{fig:lintheory} (top two panels), and compare well with data extracted from the full nonlinear dynamics (open circles) of initial fronts.
Small deviations are observed for large values of $\xi$, but are a result of the difficulty of determining a large decay length in a finite system.

It is clear from Eq.\ (\ref{eq:kstar}) that the product
\begin{equation}\label{eq:qxi}
  q\cdot\xi = \sqrt{\frac{3(3+\sqrt7)}{\sqrt7-1}} \simeq 3.2,
\end{equation}
is universal and does not depend on any wire parameters. 
This has been verified within numerical accuracy for all wire radii, as well as for two different values of $\sigma_s$ (corresponding to Au and Na).

The front velocity $v=\text{Im}[\omega(k^*)]/\text{Im}\,k^*$ is readily extracted from Eq.\ (\ref{eq:kstar}), and is found to be
\begin{equation}\label{eq:Vfront}
 v = \frac{\omega_0\sigma_s{\cal V}_a^2}{6\Ef}\sqrt{\frac{17+7\sqrt7}3}\left(-\frac{\alpha(R_0)}{\pi\sigma_s R_0}\right)^{3/2}.
\end{equation}
This result is plotted in Fig.\ \ref{fig:lintheory} (bottom panel) as a solid line, and compared with linear-front velocities obtained from the dynamical simulations (open circles). 
Agreement between the linear theory and full nonlinear dynamics is very good, showing that the front propagation is indeed governed by the linear dynamics.

Combining Eqs.\ (\ref{eq:kstar}) and (\ref{eq:Vfront}), one gets
\begin{equation}\label{eq:Vofq}
  v \propto \sigma_s q^3,
\end{equation}
so that the front velocity can be determined from the pattern wavelength. 
This relation also holds within numerical accuracy in simulations for Au and Na.

Finally, as $\alpha(R)\propto\pm(R-R_c)$ at the stability boundaries, where $\alpha(R_c)=0$, Eqs.\ (\ref{eq:kstar}) and (\ref{eq:Vfront}) provide the scaling of $v$, $q$, and $\xi$ as $|R-R_c|\rightarrow 0$, which are
\begin{equation}
  v \sim |R-R_c|^{3/2},
\end{equation}
\begin{equation}
  q \sim |R-R_c|^{1/2},
\end{equation}
and
\begin{equation}
  \xi \sim |R-R_c|^{-1/2}.
\end{equation}
%

%Their respective exponents are $3/2$, $1/2$ and $-1/2$ as $|R-R_c|\rightarrow 0$.

In cases where nonlinear effects take over the dynamics behind the front, the marginally stable front satisfying Eqs.\ (\ref{eq:kstar}--\ref{eq:Vofq}) forms and propagates for a while, but at some point it is ``invaded'' by a faster front.
This ``invading'' front has a larger $q$ and a smaller $\xi$, and satisfies neither Eq.\ (\ref{eq:qxi}), nor Eq.\ (\ref{eq:Vofq}).
This is consistant with the nonlinear marginal-stability mechanism discussed by van Saarloos \cite{vansaarloos89} where, for some range of parameters, the front becomes unstable to an ``invasion mode.''
In this case, the evolution depends on the full nonlinear dynamics, Eqs. (\ref{eq:diffusion}--\ref{eq:mu}), and the front is ``pushed'' by the invading mode, rather than ``pulled'' by the exponentially growing instability \cite{vansaarloos03}.
The new front speed, wavevector and decay length seem consistent with the analysis of Ref.\ \onlinecite{vansaarloos89}.
Although the cause of the instability of the front to the invasive mode is not clear, it seems to be related to the existence of a ``quasi-stable'' wire, i.e. a wire for which $\alpha(R)$, though still negative, is relatively small.

%%%%%%%%%%%%%%%%%%%%%%%%%%%%%%%%%%%%%%%%%%%%%%%%%%%%%%%%%%%%%%%%%%%%%%%
\section{Discussion and conclusions}\label{sec:concl}
%%%%%%%%%%%%%%%%%%%%%%%%%%%%%%%%%%%%%%%%%%%%%%%%%%%%%%%%%%%%%%%%%%%%%%%

Experimental verification of the dynamics of an unstable wire may be, at least in part, possible. 
An unstable cylinder can be prepared using a potential bias: As the magic radius intervals vary with a bias applied along the wire \cite{zhang05}, a stable cylindrical wire can be prepared at a finite bias $V$, and its stability modified by suddenly switching the bias off. 
If the bias and wire radius are chosen appropriately, the wire becomes unstable, and its dynamics can be observed using, for example, transmission electron microscopy \cite{kondo97}.
An alternative way of obtaining the same result is to stretch the wire abruptly, so that it deforms elastically into an unstable cylinder.
In both cases, the connection of the wire to macroscopic leads acts as a localized perturbation at the end of the wire triggering the instability propagation.

Actual observation of the front propagation would, however, be difficult as wire imperfections are likely to trigger the instability at several places along the wire simultaneously.
In addition, the experiment needs to be conducted at room temperature for the metal to be soft enough to allow surface diffusion, so that thermal fluctuations are likely to have the same effect.
Furthermore, the predicted radius oscillations are of the same size as the atomic granularity, and the two may thus be difficult to distinguish.

The evolution of unstable wires, discussed in Sec.\ \ref{sec:instabprop}, provides an interesting system where ideas developed in the context of front propagation into an unstable state \cite{vansaarloos88, vansaarloos89,vansaarloos03} can be successfully tested, as shown in Sec.\ \ref{sec:frontprop}.
The instability is found to grow exponentially ahead of a well-defined front, leaving a coherent pattern behind it.
In most cases, the front is found to be pulled by the instability growth ahead of it.
Its propagation is thus governed by the linearized evolution equation (\ref{eq:LinearEq}).
Simple expressions for the front velocity $v$ and decay length $\xi$, as well as the pattern wavevector $q$ have been derived, and found to be in good agreement with numerical simulations using the full non-linear dynamics.

In other cases, the linear evolution equation\ (\ref{eq:LinearEq}) fails to explain the full front dynamics.
The pulled front is found to be invaded by a faster instability that pushes the front at its higher velocity.
In this case, the front parameters depend on the full non-linear dynamics.
The presence of an invading mode seems to correspond to quasi-stable wires, for which $\alpha(R_0)$ is still negative, but small.

\section*{Acknowledgments}

I am grateful to Profs. Charles Stafford and Ray Goldstein for enlightening discussions on various aspects of this research. 
This work was supported by NSF grants 0312028 	% Charles' grant
and 0351964.					% Dan's grant

\bibliography{refs}

\begin{thebibliography}{32}
\bibitem[{Sur()}]{SurfAtoms}
\bibinfo{howpublished}{For example, more than $60\%$ of the atoms of a wire
  with a diameter of 6 atomic layers, corresponding to a conductance around
  $30\,G_0$, are at the wire's surface.}
\bibitem[{inc()}]{incompFluid}
\bibinfo{howpublished}{The incompressibility assumption is not critical.
  Alternative constraints have been considered \cite{urban04}, but do not
  significantly modify the results.}
\bibitem[{thr()}]{threshold}
\bibinfo{howpublished}{The front position thus defined is somewhat arbitrary,
  but results for the front velocity are independent of the choice of $c$
  \cite{vansaarloos03}.}
\end{thebibliography}

\end{document}